\documentclass[a4paper,10pt]{revtex4}
\usepackage{mathrsfs}
\usepackage{graphicx}
\usepackage{latexsym}
\usepackage{amsmath}
\usepackage{amssymb}
\usepackage{textcomp}
\usepackage{amsbsy}
\usepackage{graphics}
\usepackage{epstopdf}
\usepackage{color}

\begin{document}
\tolerance=5000

\title{Bottom-up reconstruction of non-singular bounce in F(R) gravity from observational indices}
\author{S.~D.~Odintsov,$^{1,2,5}$\,\thanks{odintsov@ieec.uab.es}
V.~K.~Oikonomou,$^{3,4,5}$\,\thanks{v.k.oikonomou1979@gmail.com}
Tanmoy~Paul$^{6,7}$\thanks{pul.tnmy9@gmail.com}} \affiliation{ $^{1)}$ ICREA, Passeig Luis Companys, 23, 08010 Barcelona, Spain\\
$^{2)}$ Institute of Space Sciences (IEEC-CSIC) C. Can Magrans
s/n,
08193 Barcelona, Spain\\
$^{3)}$ Department of Physics, Aristotle University of
Thessaloniki, Thessaloniki 54124,
Greece\\
$^{4)}$ International Laboratory for Theoretical Cosmology, Tomsk
State University of Control Systems
and Radioelectronics (TUSUR), 634050 Tomsk, Russia\\
$^{5)}$ Tomsk State Pedagogical University, 634061 Tomsk, Russia\\
$^{6)}$ Department of Physics, Chandernagore College, Hooghly - 712 136.\\
$^{(7)}$ Department of Theoretical Physics,\\
Indian Association for the Cultivation of Science,\\
2A $\&$ 2B Raja S.C. Mullick Road,\\
Kolkata - 700 032, India }

\tolerance=5000

\begin{abstract}
We apply the bottom-up reconstruction technique in the
context of bouncing cosmology in F(R) gravity, where the starting
point is a suitable ansatz of observable quantity (like spectral
index or tensor to scalar ratio) rather than a priori form of
Hubble parameter. In inflationary scenario, the slow roll
conditions are assumed to hold true, and thus the observational
indices have general expressions in terms of the slow-roll
parameters, as for example the tensor to scalar ratio in F(R)
inflation can be expressed as $r = 48\epsilon_F^2$ with
$\epsilon_F = -\frac{1}{H_F^2}\frac{dH_F}{dt_F}$ and $H_F$, $t_F$ are the Hubble
parameter, cosmic time respectively. However, in the bouncing cosmology (say in F(R) gravity
theory), the slow-roll conditions are not satisfied, in general,
and thus the observable quantities do not have any general
expressions that will hold true irrespective of the form of F(R).
Thus, in order to apply the bottom-up reconstruction procedure in
F(R) bouncing model, we use the conformal correspondence between
F(R) and scalar-tensor model where the conformal factor in the
present context is chosen in a way such that it leads to an
inflationary scenario in the scalar-tensor frame. Due to the
reason that the scalar and tensor perturbations remain invariant
under conformal transformation, the observable viability of the
scalar-tensor inflationary model confirms the viability of the
conformally connected F(R) bouncing model. Motivated by these
arguments, here we construct a viable non-singular bounce in F(R)
gravity directly from the observable indices of the corresponding
scalar-tensor inflationary model.
\end{abstract}

\maketitle

\section{Introduction}

The current observations indicate, with no doubt, that the present
Universe is expanding in an accelerating way. Its expansion rate
is quantified by the evolution of the Hubble parameter $H =
\dot{a}/a$, where $a(t)$ is the scale factor of the Universe. So,
when we go back in time, there are two possibilities, firstly that
the scale factor reaches the value zero and therefore the
Kretschmann scalar diverges at the time when the scale factor
becomes zero. This indicates a spacetime curvature singularity
known as Big-Bang singularity. It is a common thought that the yet
to be found quantum theory of gravity may have a significant role
in removing the Big-Bang singularity, just as happens in quantum
electrodynamics where the quantum corrections remove the classical
divergence of the Coulomb potential. The second possibility
however to describe the early-time era, that overrides the quantum
gravity era assumption, is the bouncing cosmology description
\cite{Brandenberger:2012zb,Brandenberger:2016vhg,Battefeld:2014uga,Novello:2008ra,Cai:2014bea,deHaro:2015wda,Lehners:2011kr,Lehners:2008vx,
Cheung:2016wik,Cai:2016hea,Cattoen:2005dx,Li:2014era,Brizuela:2009nk,Cai:2013kja,Quintin:2014oea,Cai:2013vm,Poplawski:2011jz,
Koehn:2015vvy,Pinto-Neto:2020xmb,Nojiri:2016ygo,Odintsov:2015ynk,Koehn:2013upa,Battarra:2014kga,Martin:2001ue,Khoury:2001wf,
Buchbinder:2007ad,Brown:2004cs,Hackworth:2004xb,Nojiri:2006ww,Johnson:2011aa,Peter:2002cn,Gasperini:2003pb,Creminelli:2004jg,Lehners:2015mra,
Mielczarek:2010ga,Lehners:2013cka,Cai:2014xxa,Cai:2007qw,Cai:2010zma,Avelino:2012ue,Barrow:2004ad,Cai:2016thi,Cai:2017tku,Wan:2015hya,
Haro:2015zda,Elizalde:2014uba,Das:2017jrl,Bari:2019yvk} where the scale factor never
becomes zero and thus the spacetime singularity is absent. In the
case of bouncing scenario, the Universe starts from a contracting
era, then it bounces off when it reaches a minimum size of the
scale factor, and starts to expand again. Thereby, the bounce
occurs at the time when $H = 0$ and $\dot{H} > 0$. Moreover,
bounce cosmology is also appealing since it can be obtained as a
cosmological solution of the theory of Loop Quantum Cosmology
\cite{Laguna:2006wr,Corichi:2007am,Bojowald:2008pu,Singh:2006im,
Date:2004fj,deHaro:2012xj,Cianfrani:2010ji,Cai:2014zga,
Mielczarek:2008zz,Mielczarek:2008zv,Diener:2014mia,Haro:2015oqa,
Zhang:2011qq,Zhang:2011vi,Cai:2014jla,WilsonEwing:2012pu}.\\
Among the non-singular bouncing models proposed so far, the matter
bounce scenario
\cite{deHaro:2015wda,Cai:2008qw,Finelli:2001sr,Quintin:2014oea,Cai:2011ci,
Haro:2015zta,Cai:2011zx,Cai:2013kja,
Haro:2014wha,Brandenberger:2009yt,deHaro:2014kxa,Odintsov:2014gea,
Qiu:2010ch,Bamba:2012ka,deHaro:2012xj,
WilsonEwing:2012pu,Cai:2011tc,Nojiri:2019lqw,Elizalde:2019tee,Elizalde:2020zcb}
gained a lot of attention because of the fact that the Universe
evolves in a way similar to a matter dominated epoch even at late
times in this scenario. However the matter bounce scenario in a
scalar-tensor theory has some problematic implications, like the
fact that the scalar power spectrum is scale invariant, so the
scalar spectral index is exactly equal to one, and the
corresponding running of the index becomes zero, which is not
compatible with Planck 2018 observations, and also the amplitude
of the tensor and scalar perturbations are of the same order,
which in turn makes the tensor-to-scalar ratio to be of the order
of unity, so it is incompatible too with the Planck constraints.
Moreover, the scalar and tensor perturbations are not stable. Here
it may be mentioned that such problems in scalar-tensor theory can
not be even resolved in a standard $F(R)$ model, because a
scalar-tensor model can be thought as an equivalent dual theory of
a $F(R)$ model, connected by a conformal transformation of the
metric (it may be mentioned that the duality between scalar-tensor and $F(R)$ model can be used to solve the F(R)
gravitational equation of motion i.e one can
solve the scalar-tensor equation of motion which are relatively easier to solve and then transform back the solutions into the corresponding
F(R) model by inverse conformal transformation, see \cite{Elizalde:2018rmz,Das:2017htt,Banerjee:2017lxi,Chakraborty:2016ydo}).
However these problems can be resolved in a Lagrange multiplier
$F(R)$ gravity model, which clearly indicates the importance of
Lagrange multiplier term in making the observable indices of a
matter bounce scenario compatible with Planck constraints
\cite{Nojiri:2019lqw}. But the energy conditions are violated near
the bouncing era (like in most of the bouncing models) in a
Lagrange multiplier $F(R)$ matter bounce model. It is the holonomy
improved Lagrange multiplier $F(R)$ gravity model which rescues
the energy condition and also makes the observable quantities
compatible with Planck results \cite{Elizalde:2019tee}. Actually
in the holonomy corrected model, the Hubble squared parameter is
proportional to quadratic and to linear powers of the effective
energy density ($\rho_{eff}$), unlike to the usual Friedmann case
where $H^2$ is proportional only to the linear power of
$\rho_{eff}$. This difference in the field equations becomes
significant near the bouncing point era and helps to rescue the
energy conditions.\\
In the earlier literature of bouncing cosmology, the scale factor or equivalently
the Hubble parameter was assumed to have an a priori specific form (maybe it is matter
bounce or quasi-matter bounce or some other models) and then
 the observational quantities were determined in a specific background
theory. However, in the present paper, we use a different approach
to study the bouncing cosmology. In particular, we use a bottom-up
reconstruction technique for non-singular bounce in an $F(R)$
gravity model, in which the observable indices are assumed to have
a specifically chosen form. Such a bottom-up approach has been
also used earlier, however in the context of F(R) inflationary
cosmology. In the case of inflation, the slow-roll conditions hold
true and thus the observable quantities can be, in general,
expressed in terms of the slow-roll parameters, as for example the
tensor to scalar ratio in F(R) inflation has a general expression
like $r = 48\epsilon_F^2$ where $\epsilon_F = -\frac{1}{H_F^2}\frac{dH_F}{dt_F}$ (during the inflationary epoch
$\epsilon_F$ remains less than unity and moreover $\epsilon_F = 1$
indicates the exit of inflation) and $H_F$, $t_F$ are the Hubble
parameter, cosmic time respectively. The authors of \cite{Odintsov:2017fnc} used this
slow-roll expression of $r$ to construct a viable F(R)
inflationary model from bottom-up reconstruction procedure.
However in the bouncing cosmology in a specific theory, say in
F(R) gravity, the scenario is different, in particular the
slow-roll conditions do not in general hold true and hence the
observable indices do not have general expressions that will hold
for any form of F(R). Thus in order to incorporate the bottom-up
reconstruction technique in the F(R) bouncing model, one may use
the conformal correspondence between F(R) and scalar-tensor model,
where the conformal factor should be chosen in such a way that it
leads to an inflationary scenario in the scalar-tensor frame. This
type of conformal equivalence between bounce and inflation has
been demonstrated in \cite{Nandi:2020sif,Odintsov:2015ynk}.
Moreover as shown in \cite{Nandi:2020sif}, the scalar and tensor
perturbations remain invariant under conformal transformation and
thus the observable viability of the scalar tensor inflationary
scenario confirms the viability of the conformally connected F(R)
bouncing scenario. Motivated by these arguments, in the present
paper, we construct a viable non-singular bounce in F(R) gravity
theory directly from the observable indices of the corresponding
scalar-tensor inflationary frame. The ansatz of the tensor to
scalar ratio we will consider for the scalar-tensor frame provides
an inflationary era
which also has an exit at a finite time.\\
The paper is organized as follows : after discussing some
essential features of $F(R)$ gravity in Sec.[\ref{sec_F(R)}], we will describe the bouncing cosmological perturbation in terms of generation era
of the perturbation modes in Sec.[\ref{sec perturbation}]. Then we will reveal the bottom-up reconstruction method in F(R) bouncing cosmology
in Sec.[\ref{sec bottomup}]. The conclusions follow in the end of the paper.

\section{Essential features of $F(R)$ gravity}\label{sec_F(R)}

Let us briefly recall some basic features of $F(R)$ gravity, which are necessary for our presentation, for reviews on
this topic see \cite{Nojiri:2010wj,Nojiri:2017ncd,delaCruzDombriz:2012xy}. The gravitational action of $F(R)$ gravity in vacuum is equal to,
\begin{eqnarray}
 S = \frac{1}{2\kappa^2} \int d^4x \sqrt{-g} F(R)
 \label{basic1}
\end{eqnarray}
where $\kappa^2$ stands for $\kappa^2 = 8\pi G = 1/M_p^2$ and also $M_p$ is the reduced Planck mass. By using the metric formalism, we vary the
action with respect to the metric tensor $g_{\mu\nu}$, and the gravitational equations read,
\begin{eqnarray}
 F'(R)R_{\mu\nu} - \frac{1}{2}F(R)g_{\mu\nu} - \nabla_{\mu}\nabla_{\nu}F'(R) + g_{\mu\nu}\Box F'(R) = 0
 \label{basic2}
\end{eqnarray}
where $R_{\mu\nu}$ is the Ricci tensor constructed from $g_{\mu\nu}$. Since the present article is devoted to cosmological context, in particular,
to non-singular bouncing cosmology, the background metric of the Universe will be assumed to be a flat Friedmann-Robertson-Walker (FRW) metric,
\begin{eqnarray}
 ds^2 = -dt_F^2 + a_F^2(t_F)\big[dx^2 + dy^2 + dz^2\big]
 \label{basic3}
\end{eqnarray}
with $t_F$ is the cosmic time and $a_F(t_F)$ being the scale factor of the Universe. For this
metric, the temporal and spatial components of Eq.(\ref{basic2})
become,
\begin{eqnarray}
0&=&-\frac{F(R)}{2} + 3\bigg(H_F^2 + \frac{dH_F}{dt_F}\bigg)F'(R) - 18\bigg(4H_F^2\frac{dH_F}{dt_F}_F + H_F\frac{d^2H_F}{dt_F^2}\bigg)F''(R)\nonumber\\
0&=&\frac{F(R)}{2} - \bigg(3H_F^2 + \frac{dH_F}{dt_F}\bigg)F'(R) + 6\bigg(8H_F^2\frac{dH_F}{dt_F} + 4\big(\frac{dH_F}{dt_F}\big)^2 
+ 6H_F\frac{d^2H_F}{dt_F^2} + \frac{d^3H_F}{dt_\mathrm{F}^3}\bigg)F''(R)\nonumber\\ 
&+&36\bigg(4H_F\frac{dH_F}{dt_F} + \frac{d^2H_F}{dt_\mathrm{F}^2}\bigg)^2F'''(R)\nonumber\\
\label{basic4}
\end{eqnarray}
respectively, where $H_F = \frac{1}{a_F}\frac{da_F}{dt_F}$ is the Hubble parameter of the
Universe. Comparing the above equations with usual Friedmann
equations, it is easy to understand that $F(R)$ gravity provides a
contribution in the energy-momentum tensor, with its effective
energy density ($\rho_{eff}$) and pressure ($p_{eff}$) given by,
\begin{eqnarray}
 \rho_{eff}&=&\frac{1}{\kappa^2}\bigg[-\frac{f(R)}{2} + 3\bigg(H_F^2 + \frac{dH_F}{dt_F}\bigg)f'(R) 
 - 18\bigg(4H_F^2\frac{dH_F}{dt_F} + H_F\frac{d^2H_F}{dt_F^2}\bigg)f''(R)\bigg]\nonumber\\
 p_{eff}&=&\frac{1}{\kappa^2}\bigg[\frac{f(R)}{2} - \bigg(3H_F^2 + \frac{dH_F}{dt_F}\bigg)f'(R) + 6\bigg(8H_F^2\frac{dH_F}{dt_F} 
 + 4\big(\frac{dH_F}{dt_F}\big)^2 + 6H_F\frac{d^2H_F}{dt_F^2}
+ \frac{d^3H_F}{dt_\mathrm{F}^3}\bigg)f''(R)\nonumber\\ 
&+&36\bigg(4H_F\frac{dH_F}{dt_F} + \frac{d^2H_F}{dt_\mathrm{F}^2}\bigg)^2f'''(R)\bigg]
\label{effective_ed}
\end{eqnarray}
respectively, where $f(R)$ is the deviation of $F(R)$ gravity from
the Einstein gravity, that is $F(R) = R + f(R)$. Thus, the
effective energy-momentum tensor (EMT) depends on the form of
$F(R)$, as expected.

\section{Cosmological perturbation : An attempt for a general expression of tensor-to-scalar ratio in F(R) bouncing scenario}\label{sec perturbation}

\begin{figure}[!h]
\begin{center}
 \centering
 \includegraphics[width=3.0in,height=2.0in]{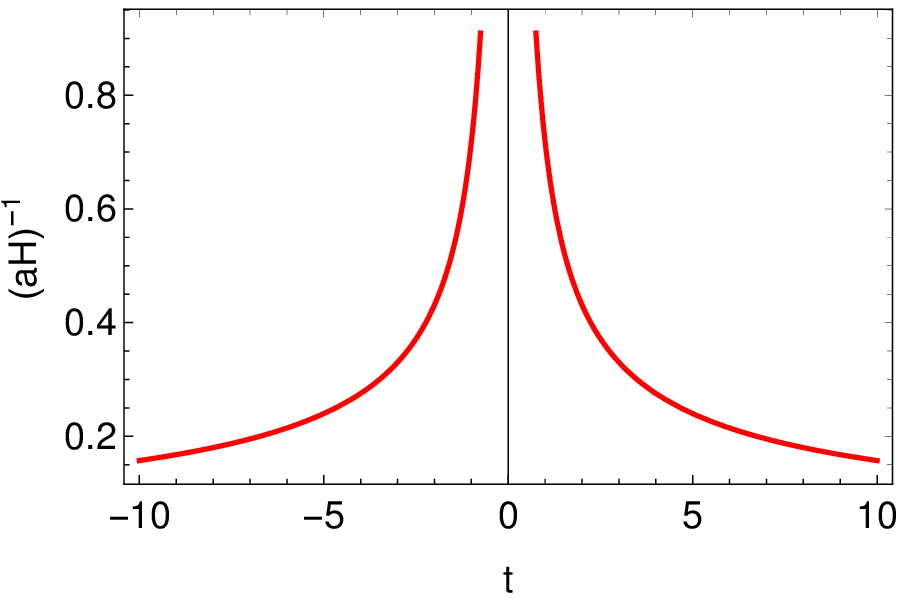}
 \includegraphics[width=3.0in,height=2.0in]{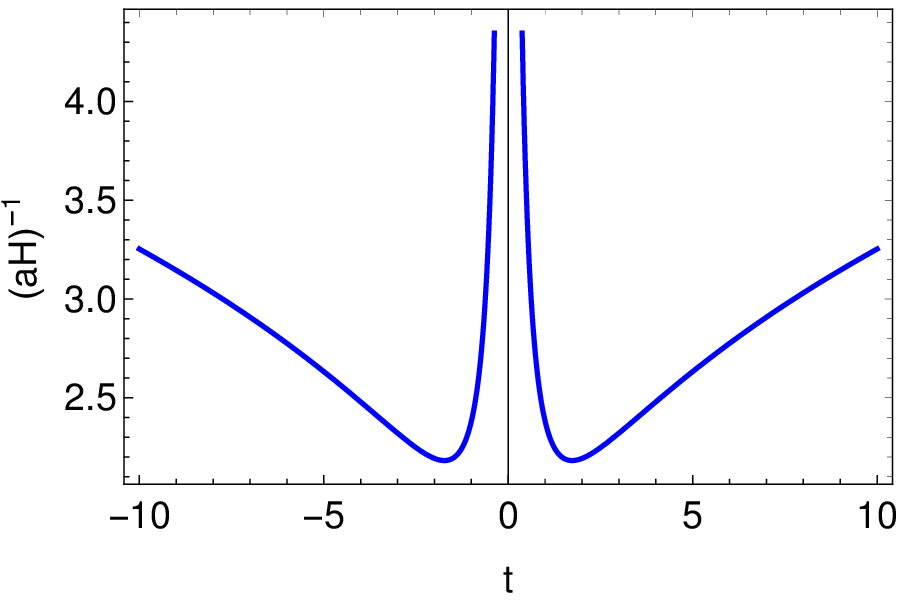}
 \caption{$Left~plot$ :The Hubble radius $r_h = \frac{1}{a_FH_F}$ as a function of the cosmic time $t_F$ for $a_F(t_F) = (t_F^2 + 1)^{4/5}$,
 where the Hubble radius decreases monotonically at both sides of the bounce and
shrinks to zero asymptotically. $Right~plot$ :
The Hubble radius $r_h = \frac{1}{a_FH_F}$ as a function of the cosmic time $t_F$ for $a_F(t_F) = (t_F^2 +1)^{1/3}$ where the Hubble radius diverges asymptotically.}
 \label{plot_horizon}
\end{center}
\end{figure}

The Universe's evolution in a general bouncing cosmology, consists
of two eras, an era of contraction and an era of expansion. Some
of the well known scale factor which correspond to a non-singular
bounce, have of the form, $a_F(t_F) = e^{\alpha t_F^2}$, $a_F(t_F)
= \cosh{t_F}$, $a_F(t_F) = (a_0t_F^2 +1)^n$, $a_F(t_F) =
e^{\frac{1}{\alpha+1}(t_F - \beta)^{\alpha+1}}$ and so on. At the
bouncing point, the Hubble parameter becomes zero and thus the
comoving Hubble radius, defined by $r_h = \frac{1}{a_FH_F}$,
diverges at the bouncing point, in all of the aforementioned
models. However the asymptotic behavior of the comoving Hubble
radius makes a difference in the above bouncing models,
specifically for some bouncing scale factors like $a_F(t_F) =
e^{\alpha t_F^2}$, $a_F(t_F) = \cosh{t_F}$, $a_F(t_F) = (a_0t_F^2
+1)^n$ for $n > 1/2$, the Hubble radius decreases monotonically at
both sides of the bounce and finally shrinks to zero size
asymptotically (see the left plot of Fig. [\ref{plot_horizon}]),
which corresponds to an accelerating late time Universe. Therefore
in such cases, the Hubble horizon goes to zero at large values of
the cosmic time, and only for cosmic times near the bouncing point
the Hubble horizon has an infinite size. So the primordial
perturbation modes relevant for present time era are generated for
cosmic times near the bouncing point, because only at that time
all the primordial modes are contained in the horizon. As the
horizon shrinks, the modes exit the horizon and become relevant
for present time observations. On other hand, some bouncing models
scale factor, like for example $a_F(t_F) = \ln{(t_F^2 + t_0^2)}$
(with $t_0$ being a constant arbitrary time), $a_F(t_F) =
(a_0t_F^2 +1)^n$ for $n < 1/2$ - lead to a divergent Hubble radius
asymptotically (see the right plot of Fig. [\ref{plot_horizon}]),
which corresponds to a decelerating Universe (due to the fact that
the Hubble radius increases) at late time. In such cases, the
perturbation modes are generated at very large negative cosmic
times, corresponding to the low curvature regime of the
contracting era, unlike to the previous situations, where the
perturbation modes are generated near the bouncing era. More
explicitly, in the latter case, the comoving wave number $k$
begins its propagation through spacetime at large negative cosmic
times, in the contracting phase on sub-Hubble scales, and exits
the Hubble radius during this phase , and re-enters the Hubble
radius during the low-curvature regime in expanding phase at the
time $t_h(k)$ (the exit and entry time are symmetric about the
bouncing point as the scale factor is itself symmetric) thus being
relevant for present time observations.
Therefore, the physical picture in the two cases is very different
with regard to when the perturbation modes are generated. However,
in both cases, the comoving curvature perturbation has to be
evolved from the contracting phase to the expanding one, followed
by the bouncing phase, in order to get the power spectrum at later
times. In the large scale limit (i.e in the super-Hubble scale $k
\ll a_FH_F$) of the contracting phase, the comoving curvature
perturbation ($\Re(k,\eta)$) satisfies the cosmological
perturbation equation
\begin{eqnarray}
 v''(k,\eta) - \frac{z''(\eta)}{z(\eta)}v(k,\eta) = 0
 \label{evolution1}
\end{eqnarray}
where $\eta$ is the conformal time defined as $dt_F = a_F(t_F)d\eta$ and
prime denotes the differentiation with respect to $\eta$
throughout the paper. The above equation is written in terms of
the canonical variable : $v(k,\eta) = z\Re(k,\eta)$, and the
variable $z(\eta)$ depends on the specific model. Since in the present context, we are interested
in F(R) model, the variable $z(\eta)$ has the form :
\begin{eqnarray}
 z(\eta(t_F)) = \frac{a_F(t_F)}{\kappa\bigg(H_F(t_F) + \frac{F''(R)}{2F'(R)}~\frac{dR}{dt_F}\bigg)} 
 \sqrt{\frac{3\big(F''(R)\big)^2}{2F'(R)}~\bigg(\frac{dR}{dt_F}\bigg)^2}
 \nonumber
\end{eqnarray}
However in terms of a general $z(\eta)$, the solution of Eq.(\ref{evolution1}) is given by,
\begin{eqnarray}
 v_c(k,\eta) = z(\eta)\bigg[D_{c}(k) + S_c(k)\int^{\eta} \frac{d\eta}{z^2}\bigg]
 \label{evolution2}
\end{eqnarray}
where the suffix 'c' denotes the contracting phase and $D_c(k)$,
$S_c(k)$ are independent of time and carry the information about
the spectra of the two modes. The above solution of $v(k,\tau)$
immediately leads to the curvature perturbation in the super-Hubble scale of the contracting phase as,
\begin{eqnarray}
 \Re_c(k,\eta) = \frac{v(k,\eta)}{z(\eta)} = D_{c}(k) + S_c(k)\int^{\eta} \frac{d\eta}{z^2}
 \label{evolution3}
\end{eqnarray}
As is evident from the above expression, the $D$ mode is a
constant mode and generally the $S$ mode behaves as an increasing
mode. Similarly in the large scale limit of the expanding phase,
the curvature perturbation has the following solution,
\begin{eqnarray}
 \Re_e(k,\eta) = D_{e}(k) + S_e(k)\int^{\eta} \frac{d\eta}{z^2}
 \label{evolution4}
\end{eqnarray}
The $D_e$ mode of the curvature perturbation is constant in time,
as is the $D_c$ mode in the contracting phase. However generally the role of
the $S$ mode becomes very different. In the expanding phase $S_e$ is
the sub-dominant decreasing mode, whereas in the contracting
phase, the opposite situation occurs, that is, the $S_c$ mode is the decreasing
mode. Therefore, the dominant mode of the curvature perturbations
in the period of expansion is $D_e$. This leads to the following power
spectrum of the curvature perturbations at late times (which is useful for the present observation),
\begin{eqnarray}
 P_e(k) = \frac{k^3}{2\pi^2} |D_e(k)|^2
 \label{evolution5}
\end{eqnarray}
Similarly the power spectrum for the tensor perturbation is given
by $P^{(T)}_e(k) = \frac{k^3}{2\pi^2} |D_e^{(T)}(k)|^2$. Moreover the curvature perturbation should be continuous and thus
$\Re_c$, $\Re_e$ have to be matched through the bouncing point as
explicitly performed in \cite{Cai:2008qw,Finelli:2001sr}. During
this matching procedure, the $D_e$ mode may inherit a contribution
from both $D_c$ and $S_c$ modes \cite{Cai:2008qw,Finelli:2001sr}. At this
stage it may be mentioned that a model will be a viable one if the
observable quantities like the spectral index ($n_s$) and the
tensor-to-scalar ratio ($r$) - defined by $n_s - 1 =
\frac{\partial\ln{P_e}} {\partial\ln{k}}\bigg|_{h.c}$ and $r =
\frac{P_e(k)}{P_e^{(T)}(k)}\bigg|_{h.c}$, are compatible with the
latest observations. Note that the subscript ``h.c'' indicates
that these must be evaluated at the horizon crossing time
instance, corresponding to a cosmic time when the perturbation
mode $k$ crosses the horizon i.e. when $k = a_FH_F$. Up to this point, we do not consider any certain form of F(R). At this stage
it deserves mentioning that the extraction of various observable quantities from the above mentioned power spectrums need
an explicit solution of the Mukhanov-Sasaki variable governed by Eq.(\ref{evolution1}) which in turn demands a $particular$ form of F(R),
in a bouncing scenario where the slow roll conditions are not valid. This spoils the generality for the
tensor to scalar ratio expression that is supposed to be true for any form of F(R). Thus
in the bouncing context, one may not be able to extract an expression for the tensor to scalar ratio which is supposed to valid
for a general F(R). On other hand, one may think about the power law parametrization where the scalar and tensor power spectrums can be
parametrized as $P_e(k) \propto \big(\frac{k}{a_FH_F}\big)^{n_s - 1}$ and $P_e^{(T)}(k) \propto \big(\frac{k}{a_FH_F}\big)^{n_T}$ respectively or
equivalently $P_e(k) = A_s\big(\frac{k}{a_FH_F}\big)^{n_s - 1}$ and $P_e^{(T)}(k) = A_T\big(\frac{k}{a_FH_F}\big)^{n_T}$ with $A_s$ being the scalar
power spectrum at the horizon crossing and thus known as scalar perturbation amplitude, for similar reason $A_T$ stands for the
tensor perturbation amplitude. However in the case of power law parametrization, the spectral index becomes independent of the
wave number $k$ and thus the running of the
spectral index vanishes i.e $\frac{dn_s}{d\ln{k}} = 0$. The vanishing $\alpha$ is not compatible with the Planck 2018 observations, which
indicates that the power law parametrization is not a viable consideration.\\

\section{Bottom-up reconstruction in F(R) bouncing cosmology}\label{sec bottomup}

The bottom-up reconstruction technique is actually motivated from the work \cite{Odintsov:2017fnc} where the bottom-up
approach is considered to check the viability of F(R) gravity in the context of slow roll inflationary scenario. The authors of
\cite{Odintsov:2017fnc} considered some specific forms, in particular, the exponential and logarithmic forms of the tensor to scalar ratio ($r$)
as a function
of the e-folding number and then compared such ansatzs of the tensor to scalar ratio with its general slow roll expression in a F(R) model i.e
with $r = 48\epsilon_F^2$ (where $\epsilon_F = -\frac{1}{H_F^2}\frac{dH_F}{dt_F}$) 
to determine the corresponding Hubble parameter in terms of e-folding number or at
a same time in terms of the cosmic time.
The determination of the Hubble parameter in turn helps to obtain the form of F(R) realizing such evolution of the
Hubble parameter from the gravitational equation of motion, which further reveals the other observable quantities like the spectral index, the
running of index etc. and consequently the model can be directly confronted with the Planck constraints. Thus
the viability of slow roll inflationary F(R) model can be judged directly from a viable ansatz of the tensor to scalar ratio rather than
starting from a Hubble parameter expression. Because the starting point is a specific form of $r = r(N)$ in such bottom-up technique,
the slow roll conditions play an important role to determine the evolution of the Hubble parameter
from the general ansatz of the tensor to scalar ratio.\\
If we want to apply the same procedure in the present context i.e
in the context of F(R) bouncing scenario, then we will be hinged
at some intermediate stage and the demonstration goes as follows :
suppose we start from some specific form of tensor to scalar ratio
which, in fact, lies within the Planck constraints for some viable
parametric regimes. According to \cite{Odintsov:2017fnc}, the next
step is to determine the Hubble parameter by comparing the ansatz
$r = r(N)$ with a general slow roll expression of the tensor to
scalar ratio, if any, valid in a F(R) bouncing scenario. However
this step is problematic, because in the case of bounce, the slow
roll conditions do not hold true in general,  and thus there is no
such general expression of the tensor to scalar ratio in a F(R)
bouncing model, unlike to the F(R) inflationary case where the
slow roll conditions are indeed true and consequently the tensor
to scalar ratio have a general expression like $r =
48\epsilon_F^2$ irrespective of the form of F(R)
\cite{Hwang:2005hb,Noh:2001ia,Hwang:2002fp}. More explicitly, the
observable quantities like the spectral index, tensor to scalar
ratio in F(R) inflationary scenario can be expressed in terms of
the slow roll parameters irrespective of the form of F(R), however
this is not the case, in general, in a bouncing scenario. Because
the slow roll conditions are not valid in a bounce model, the
bottom-up reconstruction technique considered in
\cite{Odintsov:2017fnc}
is problematic to apply in a F(R) bouncing scenario in the present context.\\

\begin{figure}[!h]
\begin{center}
 \centering
 \includegraphics[width=3.5in,height=2.0in]{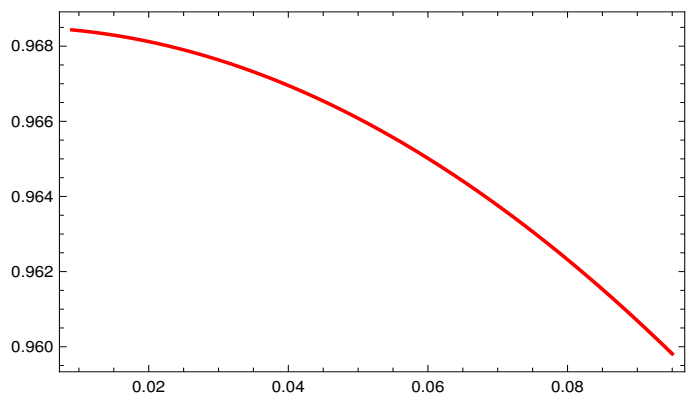}
 \caption{$n_s$ vs $\frac{\gamma}{H_0}$ for $\beta = 0.1$ and $N_T = 60$.}
 \label{plot 1}
\end{center}
\end{figure}

The above arguments clearly reveal that the validity/invalidity of the slow roll conditions is the sole reason that one can apply the
bottom-up method in an inflationary scenario but seems problematic in a bouncing model. Thus as a next attempt for applying the bottom-up reconstruction
procedure in a bouncing model, we may think the conformal correspondence of a bounce model with an inflationary one where the slow roll
conditions are indeed true. It is well known that a F(R) theory can be equivalently mapped to a scalar-tensor theory by a conformal transformation
of the spacetime metric where the scalar field potential depends on the form of F(R). Due to such conformal relation, the scale factor and the proper time
of one frame also get connected with that of the other frame. To demonstrate it briefly, let us start with a F(R) action,
\begin{eqnarray}
 S = \int d^4x \sqrt{-g}\bigg[\frac{F(R)}{2\kappa^2}\bigg]
 \label{2nd_attempt_action_fR}
\end{eqnarray}
The above action can be mapped to a scalar-tensor action by applying the following transformation of the metric
\begin{eqnarray}
 g_{\mu\nu}\longrightarrow \widetilde{g}_{\mu\nu} = e^{-\sqrt{\frac{2}{3}}~\kappa f(\phi)}~g_{\mu\nu}
 \label{conformal_transformation}
\end{eqnarray}
where $f(\phi)$ (an arbitrary function of $\phi$) is the conformal factor which is further related to the higher curvature
degrees of freedom as $F'(R) = e^{-\sqrt{\frac{2}{3}}~\kappa f(\phi)}$. If $R$ and $\widetilde{R}$ are the Ricci scalars formed by the metric
$g_{\mu\nu}$ and $\widetilde{g}_{\mu\nu}$ respectively, then they are connected by,
\begin{eqnarray}
 R = e^{-\sqrt{\frac{2}{3}}~\kappa f(\phi)}\bigg[\widetilde{R} - \kappa^2f'(\phi)^2\widetilde{g}^{\mu\nu}~\partial_{\mu}\phi\partial_{\nu}\phi
 - \sqrt{6}\kappa\widetilde{\Box}f(\phi)\bigg]
 \nonumber
\end{eqnarray}
where $\widetilde{\Box}$ is the d'Alembertian operator for $\tilde{g}_{\mu\nu}$. Using the above expression along with the aforementioned
relation between $f(\phi)$ and $F'(R)$, the following scalar-tensor action is achieved:
\begin{eqnarray}
 S = \int d^4x \sqrt{-\widetilde{g}}\bigg[\frac{\widetilde{R}}{2\kappa^2} - \frac{1}{2}\omega(\phi)\widetilde{g}^{\mu\nu}~\partial_{\mu}\phi\partial_{\nu}\phi
 - \frac{\big(A F'(A) - F(A)\big)}{F'(A)^2}\bigg]
 \label{ST action}
\end{eqnarray}
with $\omega(\phi) = f'(\phi)^2$ and
$A(x)$ being given by $F'(A) = e^{-\sqrt{\frac{2}{3}}~\kappa f(\phi)}$. Eq.(\ref{ST action}) clearly indicates that the field $\phi(x)$ acts
as a scalar field with the potential $\frac{\big(AF'(A) - F(A)\big)}{F'(A)^2} = V(A(\phi))$. Thus the higher curvature degree of freedom manifests itself
as a scalar degree of freedom with a potential $V(\phi)$ which actually depends on the form of F(R). If the F(R) model spacetime is characterized by a FRW
metric with $\eta$ be the conformal time and $a_F(\eta)$ is the scale factor i.e
\begin{eqnarray}
 ds^2 = a_F^2(\eta)\big[-d\eta^2 + \delta_{ij}dx^idx^j\big]~~~~~`,
 \label{metric_fR}
\end{eqnarray}
then the metric in the corresponding scalar-tensor model becomes
\begin{eqnarray}
 d\tilde{s}^2&=&e^{-\sqrt{\frac{2}{3}}\kappa f(\phi)}a_F^2(\eta)\big[-d\eta^2 + \delta_{ij}dx^idx^j\big]\nonumber\\
 &=&a^2(\eta)\big[-d\eta^2 + \delta_{ij}dx^idx^j\big]
 \label{metric ST}
\end{eqnarray}
with $a(\eta) = e^{-\sqrt{\frac{1}{6}}\kappa f(\phi)}a_F(\eta)$ is
the scale factor in the scalar tensor model. It may be observed
that the conformal time remains unchanged in both the frames,
however the cosmic time transforms by the way $dt =
e^{-\sqrt{\frac{1}{6}}\kappa f(\phi)}dt_F$ with $t$ being the
cosmic time in the scalar tensor theory. Before moving further, we
want to clarify the notations that we will use throughout the
paper : ($t_F$, $a_F(t_F)$) and ($t$, $a(t)$) are the cosmic time,
scale factor in the F(R) and scalar-tensor frame respectively.
Regarding the Hubble parameter, $H_F$ is reserved for the F(R)
frame while $H$ is for the scalar-tensor one. $R$ and $\widetilde{R}$ are the Ricci scalar in F(R) and scalar-tensor frame respectively. 
Moreover $\frac{d}{dt}$ is represented by an
``overdot'' (as for example $\dot{H} = \frac{dH}{dt}$), $\frac{d}{dt_F}$ is represented by itself and the other derivatives are shown by the respective 
arguments. Coming back to
Eq.(\ref{metric ST}), if $a(\eta)$ provides an inflationary
scenario in the scalar tensor frame, then by properly choosing the
conformal factor $f(\phi)$, we may get a bouncing universe in
respect to the F(R) frame scale factor $a_F(\eta)$. This type of
conformal equivalence between bounce and inflationary models has
been demonstrated in \cite{Nandi:2020sif,Odintsov:2015ynk}.
Moreover as shown in \cite{Nandi:2020sif}, the scalar and tensor
perturbations remain invariant under conformal transformation and
thus the viability of the scalar tensor inflationary scenario
confirms the viability of the conformally connected F(R) bouncing
scenario in respect to the Planck observations. Due to such
conformal connection, one may think that the viability of a F(R)
bouncing model can be investigated by looking into the
corresponding scalar-tensor inflationary frame where, due to the slow roll conditions, the bottom-up reconstruction technique can be easily applied.\\
For the purpose of applying the bottom-up reconstruction procedure in the scalar tensor model, we start with a certain ansatz of the tensor-to-scalar
ratio ($r$) which leads to an inflationary scenario. However before moving to the explicit ansatz of $r$, we present the gravitational and
scalar field equation of motion for the action (\ref{ST action}) in FRW spacetime as,
\begin{eqnarray}
 3H^2 = \kappa^2\big[\frac{1}{2}\omega\dot{\phi}^2 + V(\phi)\big]
 \label{eom1}
\end{eqnarray}
and
\begin{eqnarray}
 \omega\ddot{\phi} + \frac{1}{2}\omega'(\phi)\dot{\phi}^2 + 3H\omega\dot{\phi} + V'(\phi) = 0
 \label{eom2}
\end{eqnarray}
respectively, where the ``dot'' represents $\frac{d}{dt} =
\frac{1}{a(\eta)}\frac{d}{d\eta}$. The spatial component of the
Einstein equation i.e $2\dot{H} = -\kappa^2\omega\dot{\phi}^2$ can
be derived from the above two equations and hence is not an
independent one. Moreover the off-diagonal Einstein equations are
trivial as the off-diagonal components of the Einstein tensor
vanishes for the FRW metric. As mentioned earlier, we deal with an
inflationary scenario in the scalar-tensor (ST) model and thus the
slow roll conditions hold true in the ST frame. The slow roll
conditions are put by introducing some slow roll parameters which
is regarded to be less than unity during the inflationary period.
In the case of action (\ref{ST action}), the slow roll parameters
are defined as \cite{Nojiri:2017ncd,Hwang:2005hb},
\begin{eqnarray}
 \epsilon_1 = -\frac{\dot{H}}{H^2}~~~~~~~~,~~~~~~~~\epsilon_2 = \frac{\ddot{\phi}}{H\dot{\phi}}~~~~~~~,~~~~~~~~~\epsilon_4 = \frac{\dot{\omega}}{2H\omega}
 \label{SR conditions}
\end{eqnarray}
In a more general action like
$S = \int d^4x \sqrt{-\widetilde{g}}\bigg[\frac{1}{2\kappa^2}G(\widetilde{R},\phi)
- \frac{1}{2}\omega(\phi)\widetilde{g}^{\mu\nu}~\partial_{\mu}\phi\partial_{\nu}\phi - V(\phi)\bigg]$ (where
$G(\widetilde{R},\phi)$ is any analytic function of $\widetilde{R}$ and $\phi$), there is another slow roll parameter defined as
$\epsilon_3 = \frac{\dot{G}_R}{2HG_R}$ (with $G_R = \frac{\partial G}{\partial\widetilde{R}}$), however in the present case i.e for action
(\ref{ST action}), $G(\widetilde{R},\phi) = \widetilde{R}$ and thus the slow roll parameter $\epsilon_3$ vanishes. With the conditions
$\epsilon_i \ll 1$, the spectral index for curvature perturbation and the tensor to scalar ratio of the ST model (\ref{ST action}) are given by
\cite{Nojiri:2017ncd,Hwang:2005hb},
\begin{eqnarray}
 n_s&=&1 - 4\epsilon_1 - 2\epsilon_2 - 2\epsilon_4\nonumber\\
 r&=&8\kappa^2\bigg(\frac{\omega\dot{\phi}^2}{H^2}\bigg)
 \label{observable quantities_definitions}
\end{eqnarray}
respectively. Incorporating the gravitational equation $2\dot{H} =
-\kappa^2\omega\dot{\phi}^2$ into the above expression yields a
simplified form of the tensor to scalar ratio as follows,
\begin{eqnarray}
 r = -16\frac{\dot{H}}{H^2} = 16\epsilon_1
 \label{ratio_simplified}
\end{eqnarray}
Furthermore, the equations of motion, due to the slow roll
conditions, can be approximated as follows,
\begin{eqnarray}
 3H^2 = \kappa^2V(\phi)
 \label{SR eom1}
\end{eqnarray}
and
\begin{eqnarray}
 \frac{1}{2}\omega'(\phi)\dot{\phi}^2 + 3H\omega\dot{\phi} + V'(\phi) = 0
 \label{SR eom2}
\end{eqnarray}
Having set the stage, let us consider an ansatz of tensor-to-scalar ratio in terms of the e-folding number as,
\begin{eqnarray}
 r(N) = 16 e^{\beta\big(N(t) - N_f\big)}
 \label{ratio 2}
\end{eqnarray}
where $\beta$ is a dimensionless model parameter. 
Here it may be mentioned that the e-foldings number can be defined
as either $N(t) = \int_{t_h}^{t} Hdt$ or $N(t) =
\int_{t}^{t_\mathrm{end}} Hdt$ where $t_h$ and $t_\mathrm{end}$
are the onset and the end point of inflation respectively. Thereby
in the former case, $\dot{N} > 0$ i.e. the e-foldings number
monotonically increases with the cosmic time, while for the latter
case, the e-folding number decreases with $t$. However in the
present paper, we follow the convention for which $\dot{N} > 0$
i.e $N(t) = \int_{t_h}^{t} Hdt$. In principle, we can start with
any form of $r(N)$ i.e. any combination of functions are allowed
in the expression of the tensor to scalar ratio to start with. We
choose the particular form (\ref{ratio 2}) of $r$ in order
to proceed our calculations analytically. There may exist some
other forms of $r$ (other than (\ref{ratio 2})) for which
analytic calculations can be performed, some of them are discussed
later. The most important part is to check whether the choice of
$r$ leads to the observable compatibility with the Planck
constraints. As we now demonstrate the above choice of $r = r(N)$
leads to an inflationary cosmology in the scalar tensor frame.
Comparing Eqs.(\ref{ratio 2}) and (\ref{ratio_simplified}),
we get a first order differential equation for the Hubble
parameter as
\begin{eqnarray}
 \frac{1}{H(N)}\frac{dH}{dN} = -e^{\beta\big(N - N_f\big)}
 \label{evolution_Hubble}
\end{eqnarray}
where we use the $\frac{d}{dt} = H(N)\frac{d}{dN}$. Solving Eq.(\ref{evolution_Hubble}), we obtain
\begin{eqnarray}
 H(N) = H_0\exp{\bigg[-\frac{1}{\beta}e^{\beta(N - N_f)}\bigg]}
 \label{Hubble 2}
\end{eqnarray}
with $H_0$ is an integration constant having mass dimension [+1]. 
Such evolution of the Hubble parameter immediately leads to the
acceleration factor of the universe as $\frac{\ddot{a}}{a} =
H^2(N)\bigg(1 + \frac{1}{H(N)}\frac{dH}{dN}\bigg) = H^2(N)\bigg(1
- e^{\beta(N - N_f)}\bigg)$. Thereby the inflationary era
of the universe continues as long as the condition $1 - e^{\beta(N - N_f)} > 0$ holds, which becomes,
\begin{eqnarray}
 N < N_f (say)
 \label{acceleration period}
\end{eqnarray}
Therefore, the evolution of the Hubble parameter in
Eq.(\ref{Hubble 2}) leads to an inflationary era of the universe
and moreover the inflation has an exit at $N(t_\mathrm{end}) = N_f$ with
$t_\mathrm{end}$ is the cosmic time when the inflation ends. Thus
the total e-folding of the inflationary epoch is given by $N_T =
N(t_\mathrm{end}) - N(t_h) = \int_{t_h}^{t_\mathrm{end}} Hdt$ (the
subscript 'T' denotes the $total$ e-folding) which is considered
to be around $N_T \simeq 60$ for the CMB scale perturbation mode
having horizon crossing time is $t_h$. We will use this constraint
on $N_T$ later.\\
The de-Sitter evolution of the Hubble parameter becomes more prominent if we determine the Hubble parameter in terms of the cosmic time. Using the
relation $\dot{N} = H(N)$ along with Eq.(\ref{Hubble 2}), one can find $H = H(t)$ in the leading order of $(t-t_h)$ i.e near the onset
of inflation as,
\begin{eqnarray}
 H(t) = H_0\exp{\bigg[-\frac{1}{\beta}e^{-\beta N_f}\bigg]}\bigg\{1 - H_0(t - t_h)\exp{\bigg[-\frac{1}{\beta}e^{-\beta N_f} - \beta N_f\bigg]}\bigg\}
 \label{onset Hubble}
\end{eqnarray}
We fix the integration constant during solving $\dot{N} = H(N)$ in a way such that $N(t_h) = 0$ which is also true from the definition of
$N(t) = \int_{t_h}^{t} Hdt$ we considered. Hence the resulting evolution of $H(t)$ near the beginning of inflation (i.e $t \rightarrow t_h$) is a
quasi de-Sitter evolution. Thus as a whole, the ansatz of $r(N)$ in Eq.(\ref{ratio 2}) allows an inflationary scenario of the universe
having an exit at $N = N_f$ (or $t = t_\mathrm{end}$) and moreover the Hubble parameter evolution near the beginning of the inflation follows a quasi
de-Sitter evolution. The Hubble parameter can also be expressed in terms of the conformal time $\eta$ by using the following relation,
\begin{eqnarray}
 \eta = \int \frac{dt}{a(t)} = \int \frac{e^{-N}}{\dot{N}}~dN = \frac{1}{H_0}\int e^{-N}\exp{\bigg[-\frac{1}{\beta}e^{\beta(N - N_f)}\bigg]}~dN
 \label{conformal time}
\end{eqnarray}
The integral in the right hand side is troublesome to perform, however can be done in the limit $N \rightarrow 0$ i.e near the beginning of the
inflation and as a result, one gets
\begin{eqnarray}
 \eta(N) = -\frac{\exp{\bigg[\frac{1}{\beta}e^{-\beta N_f}\bigg]}}{H_0\bigg(1 - e^{-\beta N_f}\big)}~e^{-\big(1 - e^{-\beta N_f}\big)N}
 \label{conformal time w.r.t N}
\end{eqnarray}
We will use this expression later.
Having confirmed the inflationary scenario in the ST frame, the next task is to determine the conformal factor $f(\phi)$
(see Eq.(\ref{conformal_transformation})) in such a way that the conformally transformed F(R) frame scale factor leads to a non-singular bounce. We choose
\begin{eqnarray}
 f(\phi(N)) = \frac{\sqrt{6}}{\kappa}~\ln{\bigg[e^{-N}~\cosh{\big(\gamma \eta(N)\big)}\bigg]}
 \label{conformal factor}
\end{eqnarray}
where $\gamma$ is an arbitrary parameter for the moment and $\eta = \eta(N)$ is given in Eq.(\ref{conformal time}). Using the aforementioned
relation between $a(\eta)$ and $a_F(\eta)$ (see Eq.(\ref{metric ST})), it ie easy to see that due to the above form of
$f(\phi)$, the conformally connected F(R) frame scale factor behaves as
\begin{eqnarray}
 a_F(\eta) = \cosh{\big(\gamma \eta\big)}
 \label{FR scale factor}
\end{eqnarray}
which indeed leads to a non-singular bounce at $\eta = 0$.
Moreover near $\eta = 0$, the F(R) scale factor can be
approximated as $a_F(\eta) = 1 + \frac{\gamma^2}{2}\eta^2$ and consequently
the conformal time is related to the F(R) cosmic time by $t_F =
\int a_F(\eta)d\eta = \eta + \frac{\gamma^2\eta^3}{6} \simeq
\eta$. Thus the scale factor in terms of the cosmic time turns out
to be $a_{F}(t_F) = 1 + \frac{\gamma^2}{2}t_F^2$ from which the bouncing
behaviour (at $t_F = 0$) in the F(R) frame becomes more prominent
with respect to its cosmic time. Thereby the $f(\phi)$ in
Eq.(\ref{conformal factor}) connects an inflationary ST frame
where the Hubble parameter follows Eq.(\ref{Hubble 2}) with a F(R)
bouncing frame having the scale factor given in Eq.(\ref{FR scale
factor}). With the F(R) bouncing scale factor $a_F$, one can reconstruct the form of F(R) by using the corresponding 
Jordan frame gravitational equation of motion. For the scale factor of Eq.(\ref{FR scale factor}), the primordial perturbation modes 
generate near the bounce where the perturbation modes lie within the sub-horizon scale. 
In regard to the primordial perturbation, we will determine 
the form of F(R) near the bouncing regime. The near-bounce scale factor $a_F(t_F) = 1 + \frac{\gamma^2}{2}t_F^2$ that we have obtained immediately 
leads to the Hubble parameter and the Ricci scalar as,
\begin{eqnarray}
 H_F(t_F)&=&\frac{\gamma^2t_F}{1 + \frac{\gamma^2}{2}t_F^2} \simeq \gamma^2 t_F\nonumber\\
 R(t_F)&=&12H_F^2 + 6\frac{dH_F}{dt_F} = \frac{6\gamma^2(1 + \frac{3\gamma^2}{2}t_F^2)}{(1 + \frac{\gamma^2}{2}t_F^2)^2} \simeq 6\gamma^2 + 3\gamma^4t_F^2
 \label{rec_bounce ricci}
 \end{eqnarray}
 respectively, with the $H_F(t_F)$ and $R(t_F)$ being considered up to $O(t_F^2)$, similar to the case of the scale factor. However
 Eq.(\ref{rec_bounce ricci}) clearly indicates that the Hubble parameter varies linearly with $t_F$ and goes to zero at the bouncing point, while
 the Ricci scalar, on the other hand, becomes $R(0) = 6\gamma^2$. At a later
part, we will give an estimation of the Ricci scalar at the
bouncing point. With the above expressions, the F(R) gravitational Eq.(\ref{basic4}) becomes,
 \begin{eqnarray}
  12\gamma^2(R - 6\gamma^2)F''(R) + (R - 12\gamma^2)F'(R) + F(R) = 0\label{rec_bounce eom}
 \end{eqnarray}
Solving the above equation for $F(R)$, we get,
\begin{eqnarray}
  F(R) = \frac{6\gamma^2 D}{\sqrt{e}}R - D\sqrt{3\gamma^2\pi}~e^{-\frac{R}{12\gamma^2}}\big(R - 6\gamma^2\big)^{3/2}~
  Erfi\bigg[\frac{\sqrt{R - 6\gamma^2}}{2\sqrt{3\gamma^2}}\bigg]
  \label{rec_bounce sol2}
 \end{eqnarray}
where $Erfi[z]$ is the imaginary error function defined as $Erfi[z] = -iErf[iz]$ with $Erf[z]$ being the error function and '$i$' is the imaginary unit.
Moreover $D$ is an integration constant having mass dimension [-2]. Recall, as mentioned in 
Sec.[\ref{sec_F(R)}] that the $F(R)$ gravity contributes an effective energy-momentum tensor where the effective energy density ($\rho_{eff}$) and 
the pressure ($p_{eff}$) are given in Eq.(\ref{effective_ed}). Using these expressions of 
$\rho_{eff}$ and $p_{eff}$, it is easy to show that at the bounce 
$\rho_{eff} = \frac{1}{\kappa^2}\big[-\frac{1}{2}\big(F(R) - R\big) + 3\frac{dH_F}{dt_F}\big(F'(R) - 1\big)\big]$ and 
$\rho_{eff} + p_{eff} = \frac{1}{\kappa^2}\big[2\frac{dH_F}{dt_F}\big(F'(R) - 1\big) + 24\big(\frac{dH_F}{dt_F}\big)^2F''(R)\big]$. 
The form of F(R) as determined in Eq.(\ref{rec_bounce sol2}) leads to 
\begin{eqnarray}
 \rho_{eff} = 0~~~~~~~~~,~~~~~~~~~~\rho_{eff} + p_{eff} = -\frac{2\gamma^2}{\kappa^2}\bigg[1 + \frac{6\gamma^2D}{\sqrt{e}}\bigg]
 \nonumber
\end{eqnarray}
at $t_F = 0$. These indicate a violation of energy condition which in turn ensures a bouncing phenomena at $t_F = 0$.\\
As mentioned earlier, the observable viability of the inflationary ST
model confirms the viability of the F(R) bouncing model. Thus, in
the following, we investigate the observational viability of the 
inflationary ST frame where, recall, the Hubble parameter
has been determined directly from the observational index, in particular from the tensor-to-scalar ratio ansatz.\\
Using the slow roll field equations of the ST model (i.e Eqs.(\ref{SR eom1}) and (\ref{SR eom2})), the slow roll parameter
$\epsilon_2$ turns out to be,
\begin{eqnarray}
 \epsilon_2 = \frac{\ddot{\phi}}{H\dot{\phi}} = -\frac{\big(3\dot{H}\omega + 3H\dot{\omega} + \frac{1}{2}\ddot{\omega}\big)}
 {H\big(3H\omega + \frac{1}{2}\dot{\omega}\big)}
 \nonumber
\end{eqnarray}
where $\omega(\phi)$ is the self kinetic coupling of the scalar field, which is further related to the conformal factor as
$\omega(\phi) = f'(\phi)^2$. Plugging the above expression of $\epsilon_2$ into Eq.(\ref{observable quantities_definitions}) yields
the spectral index as follows,
\begin{eqnarray}
 n_s = 1 + \frac{4\dot{H}}{H^2} + \frac{2\big(3\dot{H}\omega + 3H\dot{\omega} + \frac{1}{2}\ddot{\omega}\big)}
 {H\big(3H\omega + \frac{1}{2}\dot{\omega}\big)} - \frac{\dot{\omega}}{\omega H}
 \label{spectral 1}
\end{eqnarray}
with, recall, the ``dot'' symbolizes $\frac{d}{dt}$ (i.e the
derivative with respect to the ST frame cosmic time). We will
determine the scalar spectral index in terms of e-folding number
and for this purpose what we need is the following identities:
\begin{eqnarray}
 \frac{d}{dt} = H(N)\frac{d}{dN}~~~~~~~~~~,~~~~~~~~~~~~\frac{d^2}{dt^2} = H^2(N)\frac{d^2}{dN^2} + H\frac{dH}{dN}\frac{d}{dN}
 \label{identity}
\end{eqnarray}
Using the above identities along with the aforementioned relation between $\omega(\phi)$ and $f(\phi)$, we determine various terms
present in the right hand side of Eq.(\ref{spectral 1}), as follows:
\begin{eqnarray}
 &\bigg[&\frac{3\dot{H}\omega + 3H\dot{\omega} + \frac{1}{2}\ddot{\omega}}{3H\omega + \frac{1}{2}\dot{\omega}}\bigg]
 = \frac{H}{\bigg[f''(N) + \big(3 - e^{\beta(N - N_f)}\big)f'(N)\bigg]}\times\bigg[-3f'(N)e^{\beta(N - N_f)}\nonumber\\
 &+&\big(f''(N) - f'(N)e^{\beta(N - N_f)}\big)\big(6 - 3e^{\beta(N - N_f)} + \frac{f''(N)}{f'(N)}\big)
 + \big(f'''(N) - f''(N)e^{\beta(N - N_f)} - \beta f'(N)e^{\beta(N - N_f)}\big)\bigg]
 \label{term 1}
\end{eqnarray}
and
\begin{eqnarray}
 \frac{\dot{\omega}}{\omega H} = \frac{2\bigg(f''(N) - f'(N)e^{\beta(N - N_f)}\bigg)}{f'(N)}
 \label{term 2}
\end{eqnarray}
where $f'(N) = \frac{df}{dN}$ (also the higher derivatives have the respective meaning) and in
determining the above expressions, we neglect the acceleration term of the scalar field due to the slow roll conditions. Recall, the conformal
factor $f(N)$ is chosen in such a way in Eq.(\ref{conformal factor}) that it leads to a non-singular bounce in the F(R) frame.
However in order to determine the explicit form of $f(N)$ we need the functional behaviour of $\eta = \eta(N)$ which in turn demands
to perform the integral of Eq.(\ref{conformal time}). As demonstrated earlier in Eq.(\ref{conformal time w.r.t N}),
this integral can be performed in the limit $N \rightarrow 0$ i.e
near the horizon crossing time, which is indeed sufficient in the present context as the observable quantities like the spectral index, tensor-to-scalar
ratio are eventually determined at the horizon crossing instance. As a result, the conformal factor in terms of the e-folding number takes the following
form:
 \begin{eqnarray}
 f(N) = \frac{\sqrt{6}}{\kappa}\bigg\{-N + \ln{\bigg[\cosh{\bigg(\frac{\gamma P(N)}{H_0(1 - e^{-\beta N_f})}\bigg)}\bigg]}\bigg\}
 \label{f}
\end{eqnarray}
with $P(N) = \exp{\big[-\big(1 - e^{-\beta N_f}\big)N + \frac{1}{\beta}e^{-\beta N_f}\big]}$. Consequently, $f'(N)$, $f''(N)$ and
$f'''(N)$ are obtained as,
\begin{eqnarray}
 f'(N) = \frac{\sqrt{6}}{\kappa}\bigg\{-1 - \frac{\gamma}{H_0}~P(N)~\tanh{\bigg(\frac{\gamma P(N)}{H_0(1 - e^{-\beta N_f})}\bigg)}\bigg\}
 \label{derivative f}
\end{eqnarray}
\begin{eqnarray}
 f''(N) = \frac{\sqrt{6}}{\kappa}\bigg\{\frac{\gamma^2}{H_0^2}~P^2(N)~\cosh^{-1}{\bigg(\frac{\gamma P(N)}{H_0(1 - e^{-\beta N_f})}\bigg)}
 + \frac{\gamma\big(1 - e^{-\beta N_f}\big)}{H_0}~P(N)~\tanh{\bigg(\frac{\gamma P(N)}{H_0(1 - e^{-\beta N_f})}\bigg)}\bigg\}
 \label{double derivative f}
\end{eqnarray}
and
\begin{eqnarray}
 f'''(N)&=&\frac{\sqrt{6}}{\kappa}\bigg\{-\frac{3\gamma^2\big(1 - e^{-\beta N_f}\big)}{H_0^2}~P^2(N)
 \cosh^{-2}{\bigg(\frac{\gamma P(N)}{H_0(1 - e^{-\beta N_f})}\bigg)}
 - \frac{\gamma\big(1 - e^{-\beta N_f}\big)^2}{H_0}P(N)\tanh{\bigg(\frac{\gamma P(N)}{H_0(1 - e^{-\beta N_f})}\bigg)}\nonumber\\
 &+&\frac{2\gamma^3}{H_0^3}~P^3(N)~
 \cosh^{-2}{\bigg(\frac{\gamma P(N)}{H_0(1 - e^{-\beta N_f})}\bigg)}\tanh{\bigg(\frac{\gamma P(N)}{H_0(1 - e^{-\beta N_f})}\bigg)}\bigg\}
 \label{triple derivative f}
\end{eqnarray}
respectively. Plugging the expressions of Eqs.(\ref{term 1}) and (\ref{term 2}) into Eq.(\ref{spectral 1}), one gets the final form of the spectral
index in terms of the e-folding number as,
\begin{eqnarray}
 &n_s&= 1 - 2e^{\beta(N - N_f)} - 2\frac{f''(N)}{f'(N)} + \frac{2}{\bigg[f''(N) + \big(3 - e^{\beta(N - N_f)}\big)f'(N)\bigg]}\times
 \bigg[-3f'(N)e^{\beta(N - N_f)}\nonumber\\
 &+&\big(f''(N) - f'(N)e^{\beta(N - N_f)}\big)\big(6 - 3e^{\beta(N - N_f)} + \frac{f''(N)}{f'(N)}\big)
 + \big(f'''(N) - f''(N)e^{\beta(N - N_f)} - \beta f'(N)e^{\beta(N - N_f)}\big)\bigg]
 \label{final n_s}
\end{eqnarray}
where $f^{(i)}(N)$ (with $i = 0,1,2,3$) are given above and recall that $t_h$ is the horizon crossing instance.
Thus Eqs.(\ref{final n_s}) and (\ref{ratio 2}) provide the final forms of the scalar spectral index
and the tensor-to-scalar ratio (as a function of e-folding number) in the ST frame respectively. With these expressions of $n_s$ and $r$,
now we can confront the model with the Planck 2018 constraints \cite{Akrami:2018odb}, which constrain the observational indices as follows,
\begin{equation}
\label{planckconstraints}
n_s = 0.9649 \pm 0.0042\, , \quad r < 0.064\, .
\end{equation}
Eq.(\ref{ratio 2}) clearly indicates that $r$ depends on $N_f -
N(t_h)$ ($= N_T$ i.e the total e-folding of the inflationary era)
and $\beta$, while from Eq.(\ref{final n_s}), it is easy to
observe that the spectral index depends on $N_T$ and the
dimensionless parameters $\beta$, $\frac{\gamma}{H_0}$. The
dependence of $n_s$ on the parameter $\gamma/H_0$ actually arises
from the conformal factor which has $\gamma$ dependent term.
Considering $N_T = 60$, the tensor to scalar ratio lies within the
Planck constraints for $\beta > 0.092$. Thus taking $\beta = 0.1$
and $N_T = 60$, the spectral index is compatible with the Planck
results if the parameter $\frac{\gamma}{H_0}$ lies within the
range given by $10^{-3} \lesssim \frac{\gamma}{H_0} \lesssim 0.1$;
this is depicted in Fig.[\ref{plot 1}]. The parameter $H_0$ is
approximately the de-Sitter Hubble parameter during inflationary
epoch (see Eq. (\ref{onset Hubble})), which is generally
considered as $H_0 \simeq 10^{16}$GeV $= 10^{-3}/\kappa$ with
$\kappa = 10^{19}$GeV. With such consideration of $H_0$ and due to
the aforementioned viable range of $\frac{\gamma}{H_0}$, the
parameter $\gamma$ lies within $\gamma =
\frac{1}{\kappa}[10^{-6},10^{-4}]$. This in turn estimates the
F(R) Ricci scalar at the bounce as
$R(\eta = 0) = 6\gamma^2 \sim [10^{26},10^{30}]$(GeV)$^2$.\\

Thus the scalar-tensor inflationary observable quantities are simultaneously compatible with the Planck constraints for the parametric
ranges given by $N_T = 60$, $\beta = 0.1$ and $10^{-3} \lesssim \frac{\gamma}{H_0} \lesssim 0.1$ respectively. Being the scalar and tensor perturbations
remain invariant under conformal transformation, the observable viability of the scalar-tensor inflationary model in turn confirms
the viability of the conformally connected F(R) bouncing model where the scale factor behaves as
$a_F(\eta) = \cosh{\big(\gamma\eta\big)}$. Thus a viable F(R) bouncing model can be constructed directly from the observable
indices of the corresponding scalar-tensor inflationary frame.\\

Before concluding we would like to mention that apart from the ansatz (\ref{ratio 2}) of the tensor to scalar ratio, some other forms of $r = r(N)$
also lead to analytic results. Some of them are given by,
\begin{eqnarray}
 r(N) = \frac{16\alpha}{\beta - N}~~~~~~~~~~~,~~~~~~~~~~~~~r(N) = \frac{1}{\beta^2}
 \label{different ansatz}
\end{eqnarray}
etc. For the former case i.e for $r(N) = \frac{16\alpha}{\beta -
N}$, the Hubble parameter comes as $H(N) = H_0\big(\beta -
N\big)^{\alpha}$, while comparing $r = -\frac{16H'(N)}{H(N)}$ with
the latter one yields $H(N) = H_0e^{-N/\beta}$. However the Hubble
parameter $H(N) = H_0\big(\beta - N\big)^{\alpha}$ vanishes at a
finite e-folding $N = \beta$ which is not physical at all. On
other hand, $H(N) = H_0e^{-N/\beta}$ leads to an ever accelerating
universe i.e the inflationary scenario of the scalar-tensor frame
gets no exit. Thus it is clear that although the ansatz of Eq.
(\ref{different ansatz}) provide analytic results, such forms of
$r=r(N)$ suffer with some severe problems, unlike the form $r(N) =
16 e^{\beta\big(N(t) - N_f\big)}$
that we have considered in the present paper in Eq.(\ref{ratio 2}) which seems to free from such problems.\\

\section{Conclusions}\label{sec_conclusion}

In the present work, we have applied the bottom-up reconstruction
technique to construct a viable non-singular bounce in F(R)
gravity, where the starting point is to consider a suitable ansatz
of observational quantities, like the scalar spectral index or
tensor to scalar ratio as function of e-foldings number, rather
than a priori form of Hubble parameter. The bottom-up procedure
can be directly applied in inflationary context where, due to the
slow roll conditions, the observable quantities can be expressed
in terms of the slow roll parameters in general. However in
bouncing case (say in F(R) gravity), the scenario is different, in
particular the slow roll conditions in a bouncing model are not
true and hence the observable indices do not have any general
expressions that will be valid for any form of F(R). Thus in order
to apply the bottom-up reconstruction technique in F(R) bouncing
model, we have used the conformal equivalence between F(R) and
scalar-tensor model, where the conformal factor is chosen in such
a way so that it leads to an inflationary era in the scalar-tensor
frame. Thereby the conformal factor bridges a F(R) non-singular
bounce model to a scalar-tensor inflationary model. Moreover the
observational viability of the scalar-tensor inflationary frame,
where the bottom-up reconstruction can be applied, confirms the
viability of the conformally connected F(R) bouncing model.
Keeping these arguments in mind, we try to construct a viable F(R)
bouncing scenario directly from the observable indices of the
corresponding scalar-tensor (ST) model, in particular we start
with a suitable ansatz of the tensor-to-scalar ratio of the ST
frame in terms of e-foldings number. The ansatz of $r = r(N)$
corresponds to an inflationary era in the scalar-tensor frame,
which also has an exit at a finite time. On other hand, due to the
suitably considered conformal factor, the F(R) frame scale factor
behaves as $a_F(\eta) = \cosh{\big(\gamma\eta\big)}$ which
indicates a non-singular bounce at $\eta = 0$. With the ansatz $r
= r(N)$ along with the conformal factor, we investigate the
viability of the ST inflationary model in respect to the Planck
constraints and as a result, the observable quantities like the
spectral index, tensor-to-scalar ratio are found to lie within the
constraints for a certain parametric ranges. This in turn confirms
the observable viability of the $F(R)$ bouncing model. Thus a
viable $F(R)$ bouncing model is constructed directly from the
tensor-to-scalar ratio ansatz of the corresponding scalar-tensor
inflationary model.

\end{document}